\documentclass{PoS}

\usepackage{graphicx}
\usepackage{aps_macros}
\usepackage[dvipsnames,svgnames]{xcolor} 
\usepackage{amsmath}
\usepackage{multirow}
\usepackage{accents}
\usepackage{url} 
\usepackage{soul} 
\usepackage{amsmath}
\usepackage{xspace}
\usepackage{lineno}

\mathchardef\mhyphen="2D
\newcommand{\vect}[1]{\vec{#1}}
\newcommand{\roughly}{\ensuremath{ {\sim}\,} }

\newlength{\dhatheight}

\newcommand{\unit}[1]{\ensuremath{\mathrm{\,#1}}\xspace}
\newcommand{\MeV}{\unit{MeV}}
\newcommand{\GeV}{\unit{GeV}}
\newcommand{\TeV}{\unit{TeV}}
\newcommand{\degree}{\ensuremath{{}^{\circ}}\xspace}

\newcommand{\cm}{\unit{cm}}

\newcommand{\GeVcmcube}{\ensuremath{\GeV\cm^{-3}}\xspace}

\newcommand{\Fermi}{\textit{Fermi}\xspace}

\newcommand{\passsevenrep}{\code{Pass\,7 Reprocessed}}

\newcommand{\passeight}{\code{Pass\,8}}

\newcommand{\like}{\ensuremath{\mathcal{L}}\xspace} 
\newcommand{\pseudolike}{ {\tilde{\like}} \xspace}   
\newcommand{\given}{\ensuremath{ \,|\, }\xspace}

\newcommand{\data}{ \ensuremath{ \mathcal{D} }\xspace } 

\newcommand{\params}{\ensuremath{\vect{\alpha}}\xspace}
\newcommand{\sig}{\ensuremath{\mu}\xspace}
\newcommand{\bkg}{\ensuremath{\theta}\xspace}
\newcommand{\interest}{\ensuremath{\vect{\sig}}\xspace}
\newcommand{\nuisance}{\ensuremath{\vect{\bkg}}\xspace}

\newcommand{\Jlike}{\ensuremath{\like_{J}}\xspace}
\newcommand{\Jsigma}{\ensuremath{\sigma_{i}}\xspace}
\newcommand{\Jtrue}{\Ji}

\newcommand{\Jobsi}{\ensuremath{J_{\rm{obs},i}}\xspace}
\newcommand{\logtenJtrue}{\logtenJi}
\newcommand{\logtenJobs}{\ensuremath{ {\log_{10}{(\Jobsi)}} }\xspace}

\newcommand{\Ji}{\ensuremath{J_i}\xspace}

\newcommand{\logtenJi}{\ensuremath{{\log_{10}{(J_i)}}}\xspace}

\newcommand{\CL}{CL\xspace}

\newcommand{\code}[1]{\texttt{#1}\xspace}

\newcommand{\DM}{\ensuremath{\mathrm{DM}}}
\newcommand{\mDM}{\ensuremath{m_\DM}\xspace}

\newcommand{\sigmav}{\ensuremath{\langle \sigma v \rangle}\xspace}

\newcommand{\bbbar}{\ensuremath{b \bar b}\xspace}

\newcommand{\tautau}{\ensuremath{\tau^{+}\tau^{-}}\xspace}

\newcommand{\Jfactor}{J-factor\xspace}
\newcommand{\Jfactors}{J-factors\xspace}

\newcommand{\NRAND}{300\xspace}

\providecommand\physrep{\ref@jnl{Phys.~Rep.}}%
\providecommand\apjs{\ref@jnl{ApJS}}%
\providecommand{\jcap}{\ref@jnl{JCAP}}%

\title{Dark Matter Searches with the Fermi-LAT in the Direction of
  Dwarf Spheroidals}

\ShortTitle{Dark Matter Searches with the Fermi-LAT in the Direction of
  Dwarf Spheroidals}

\author{\speaker{Matthew Wood}\\
  Kavli Institute for Particle Astrophysics and Cosmology, SLAC National Accelerator Laboratory\\
  E-mail: \email{mdwood@slac.stanford.edu}}
\author{Brandon Anderson\\
  Department of Physics, Stockholm University\\
  E-mail: \email{brandon.anderson@fysik.su.se}}
\author{Alex Drlica-Wagner\\
  Center for Particle Astrophysics, Fermi National Accelerator Laboratory\\
  E-mail: \email{kadrlica@fnal.gov}}
\author{Johann~Cohen-Tanugi\\
  Laboratoire Univers et Particules de Montpellier, Universit\'e Montpellier\\
  E-mail: \email{johann.cohen-tanugi@lupm.in2p3.fr}}
\author{Jan Conrad\\
  Department of Physics, Stockholm University\\
  E-mail: \email{conrad@fysik.su.se}}

\author{on behalf of the \Fermi-LAT Collaboration}

\abstract{The dwarf spheroidal satellite galaxies of the Milky Way are
  some of the most dark-matter-dominated objects known. Due to their
  proximity, high dark matter content, and lack of astrophysical
  backgrounds, dwarf spheroidal galaxies are widely considered to be
  among the most promising targets for the indirect detection of dark
  matter via gamma rays. Here we report on gamma-ray observations of
  Milky Way dwarf spheroidal satellite galaxies based on 6 years of
  \Fermi Large Area Telescope data processed with the new Pass 8
  reconstruction and event-level analysis. None of the dwarf galaxies
  are significantly detected in gamma rays, and we present upper
  limits on the dark matter annihilation cross section from a combined
  analysis of the 15 most promising dwarf galaxies. The constraints
  derived are among the strongest to date using gamma rays and lie
  below the canonical thermal relic cross section for WIMPs of mass
  $\lesssim 100\GeV$ annihilating via the \bbbar and \tautau
  channels.}

\FullConference{The 34th International Cosmic Ray Conference,\\
		30 July- 6 August, 2015\\
		The Hague, The Netherlands}

\begin{document}

\section{Introduction}

Overwhelming evidence supports the existence of dark matter (DM) based
on its gravitational influence on ordinary matter.  Although the DM
paradigm is well-established, the nature of the particle that may
constitute DM is not yet known.  One of the leading candidates for the
DM particle are weakly interacting massive particles (WIMPs) that
existed in thermal equilibrium with standard model particles in the
early universe.  The relic density of WIMPs is set by ``freeze-out''
when the rate of WIMP interactions drops below the expansion rate of
the Universe.  For particles with an annihilation cross section on the
weak interaction scale such as WIMPs, this process can naturally
produce a relic density equal to the density of DM observed today.

Self-annihilation of WIMPs in regions with high DM density can produce
stable secondary particles detectable by ground- and space-based
observatories.  The gamma rays produced in these annihilations are an
especially attractive target for experimental searches.  Unlike
charged secondaries, gamma rays can be traced back to their point of
origin enabling DM signals to be more easily discriminated from
astrophysical foregrounds.  The Large Area Telescope onboard the Fermi
gamma-ray observatory is sensitive in the energy range from 20 MeV to
more than 300 GeV and is well-suited to search for the gamma-ray
annihilation signature of DM.  With its excellent angular resolution,
the Fermi-LAT can search for signals in a variety of targets including
the Galactic Center, dwarf spheroidal galaxies (dSphs), and galaxy
clusters.

Milky Way (MW) dSphs are especially promising targets due to their proximity
and the absence of intrinsic sources of gamma-ray emission.  The MW
dSphs are also predominantly found in high-latitude regions of the sky
where the diffuse gamma-ray foregrounds are lower and more easily
discriminated from a potential DM signal.  The DM content of these
objects can be robustly measured by kinematic modeling of the stellar
velocities of member stars.  The expected DM signal is proportional to
the line-of-sight integral of the DM distribution known as the
\Jfactor.  \Jfactors for known dSphs generally fall in the range
between $10^{17} \GeV^{2} \cm^{-5}$ and $10^{20} \GeV^{2} \cm^{-5}$.
By comparison, models for the DM distribution of the Galactic Center
region predict \Jfactors at least an order of magnitude larger.
However searches for DM signals in the Galactic Center are complicated
by the systematic uncertainties associated with the diffuse
astrophysical foregrounds which are brightest in this region of the
Galaxy.  Because the astrophysical foregrounds are much lower
and more easily modeled in dSphs, DM searches in dSph galaxies can
offer comparable sensitivity to those in the Galactic Center.

In this paper we present an updated search for DM signals using the
sample of 25 dSphs from \cite{Ackermann:2013yva} based on 6 years of
LAT data processed with \passeight.  \footnote{The content of this
  proceeding is a condensed version of \cite{Ackermann:2015zua}.  See
  this paper and its supplementary materials for additional details.}
Following the approach of previous works
\cite{Ackermann:2011wa,GeringerSameth:2011iw,Mazziotta:2012ux,Ackermann:2013yva,Geringer-Sameth:2014qqa,2015arXiv150203081A},
we use a joint analysis to combine the data from multiple dSphs and
increase the sensitivity of our search.  We discuss the implications
of these measurements in the context of the Galactic center excess and
other experimental constraints on the DM annihilation cross section.

\section{Pass 8}

One of the main improvements of the present analysis with respect to
prior works is the use of the \passeight data set.  \passeight is a
comprehensive revision to the LAT event-level analysis that
reconstructs and classifies individual events from the raw data
collected by the LAT subsystems.  With respect to the current
\passsevenrep data release, \passeight contains a number of
improvements to the event reconstruction including algorithms to
identify and remove instrumental pile-up, an improved energy
reconstruction that extends the LAT energy range below 100\MeV and
above 1\TeV, and a more robust pattern-recognition algorithm for track
reconstruction~\cite{Atwood:2013rka}.

The event classification in \passeight has been fully redesigned and
provides a completely new set of event classes for high-level science
analysis.  The \passeight event class selections have been optimized
using boosted decision trees to maximize the rejection power for
cosmic-ray background events while maintaining a high efficiency for
gamma rays.  The \passeight event analysis also expands the existing
classification framework by adding \emph{event types}, subdivisions of
the event classes based on event-by-event uncertainties in the
directional and energy measurements.  The PSF event types are ordered
by the quality of the angular resolution from PSF0 (worst) to PSF3
(best).  At 3.16\GeV the 68\% (95\%) containment radii of the
incidence-angle-averaged PSF for the best and worst PSF event types
(PSF3 and PSF0) is 0.17\degree (0.35\degree) and 0.92\degree
(2.3\degree).  Our maximum likelihood analysis of the dSphs combines
the four P8R2\_SOURCE\_V6 PSF event types in a joint likelihood
function.  We estimate that splitting the event sample by PSF event
type improves the sensitivity of our analysis by \roughly 10\%.
Considering both the improvements in instrument performance from
\passeight and the expanded six-year data set, we expect an
improvement of between 1.7 and 2.2 in the median sensitivity to WIMP
annihilations as compared to the four-year analysis of
\cite{Ackermann:2013yva} based on \passsevenrep data.

\section{Analysis}


We analyzed LAT data collected during the first six years of the
mission (2008-08-04 to 2014-08-05).  From this data set we select
events in the energy range between 500\MeV and 500\GeV in the
\passeight SOURCE class.  We apply a zenith angle cut of $<
100\degree$ to remove photons from the Earth Limb.  Time periods
around bright gamma-ray bursts and solar flares are also excised
following the prescription of the \Fermi-LAT third source catalog
(3FGL)~\cite{Ackermann:2015hja}.  Data are binned into $10\degree
{\times} 10\degree$ square regions of interest (ROIs) in Galactic
coordinates centered at the position of each dSph.  We use a spatial
binning of 0.1\degree and 24 logarithmic energy bins (8 bins per
energy decade).

We perform a standard binned likelihood of the 25 dSphs in our sample
using the Fermi Science Tools and the P8R2\_SOURCE\_V6 IRFs.  We use
the standard templates for isotropic and galactic diffuse emission.
\footnote{Specifically we use the \textit{v06} release of the Galactic
  interstellar emission model and isotropic templates documented at
  \url{http://fermi.gsfc.nasa.gov/ssc/data/access/lat/BackgroundModels.html}.}
Point-sources within each ROI are taken from the 3FGL
catalog~\cite{Ackermann:2015hja}.  Each ROI is first fit with a
baseline model in which the normalizations of the diffuse components
and catalog sources within the ROI are left free to vary.  For each
dSph we use a spatial profile corresponding to an NFW DM density
profile with halo scale radii and characteristic densities taken from
\cite{Ackermann:2013yva}.

After fitting the baseline model to each ROI, we derive a set of
bin-by-bin likelihoods following the methodology of
\cite{Ackermann:2013yva}.  In each energy bin, we scan the
normalization of the DM source while leaving the parameters of the
background model fixed.  The set of bin-by-bin likelihood functions
for each target is then used to reconstruct the global likelihood for
a variety of DM spectral models.

Given the LAT likelihood function for a dSph, we construct a
likelihood for the parameters of the DM model ($\interest$) from the
product of LAT and \Jfactor likelihoods.  We define the likelihood
function for target $i$ as
\begin{equation}\label{eqn:full_likelihood}
\begin{aligned}
\pseudolike_i(\interest,\nuisance_i =\{\params_{i},J_{i}\}\given
\data_i) = &\like_i(\interest,\nuisance_i \given \data_i) 
\Jlike(\Jtrue \given \Jobsi, \Jsigma).
\end{aligned}
\end{equation}
where $\nuisance_{i}$ is the set of nuisance parameters that includes
both parameters from the LAT analysis ($\params_{i}$) and the dSph
\Jfactor (\Jtrue) and $\data_{i}$ is the gamma-ray data.  The second
term in Eq. \ref{eqn:full_likelihood} is the \Jfactor likelihood function
which we use to model the uncertainty in the \Jfactor arising from the
finite precision of the stellar kinematic analysis.  We use
a log-normal parameterization of the \Jfactor likelihood given by
\begin{equation}\label{eqn:jfactor_likelihood}
\begin{aligned}
  \Jlike(\Jtrue \given \Jobsi, \Jsigma) &= \frac{1}{\ln(10) \Jobsi
    \sqrt{2 \pi} \sigma_i} 
  \times e^{-\left(\logtenJtrue-\logtenJobs\right)^2/2\sigma_i^2},
\end{aligned}
\end{equation}
where \Jtrue is the true value of the \Jfactor and \Jobsi is the
measured \Jfactor with error \Jsigma.  We obtain the \Jfactor
likelihood parameterization by fitting a log-normal function with peak
value \Jobsi to the posterior distribution for each \Jfactor as
derived by \cite{Martinez:2013ioa}.

We perform a combined analysis for DM signals in our dSph sample by
forming a joint likelihood from the individual dSph likelihoods.  Our
combined sample consists of 15 dSphs that have kinematically
determined \Jfactors and avoid ROI overlap and corresponds to the same
set that was used in \cite{Ackermann:2013yva}.  The combined
likelihood function is given by the product of the individual dSph
likelihoods,
\begin{equation}\label{eqn:joint_likelihood}
\pseudolike(\interest,\{\nuisance_i\}\given \data) = \prod_{i}\pseudolike_i(\interest,\nuisance_i\given \data_i),
\end{equation}
where the index $i$ runs over the 15 dSphs in our combined sample.
With the combined dSph likelihood, we test the hypothesis of a DM
signal in the entire sample where the expected signal in each dSph is
weighted in proportion to its \Jfactor.  For a given mass and
annihilation channel, upper limits on the annihilation cross section
are evaluated with the delta-log-likelihood technique, requiring a
change in the profile log-likelihood of 2.71/2 from its maximum for a
95\% \CL upper limit \cite{Rolke:2004mj}.

\section{Results}

\begin{figure*}[ht]
  \centering
  \includegraphics[width=0.49\textwidth]{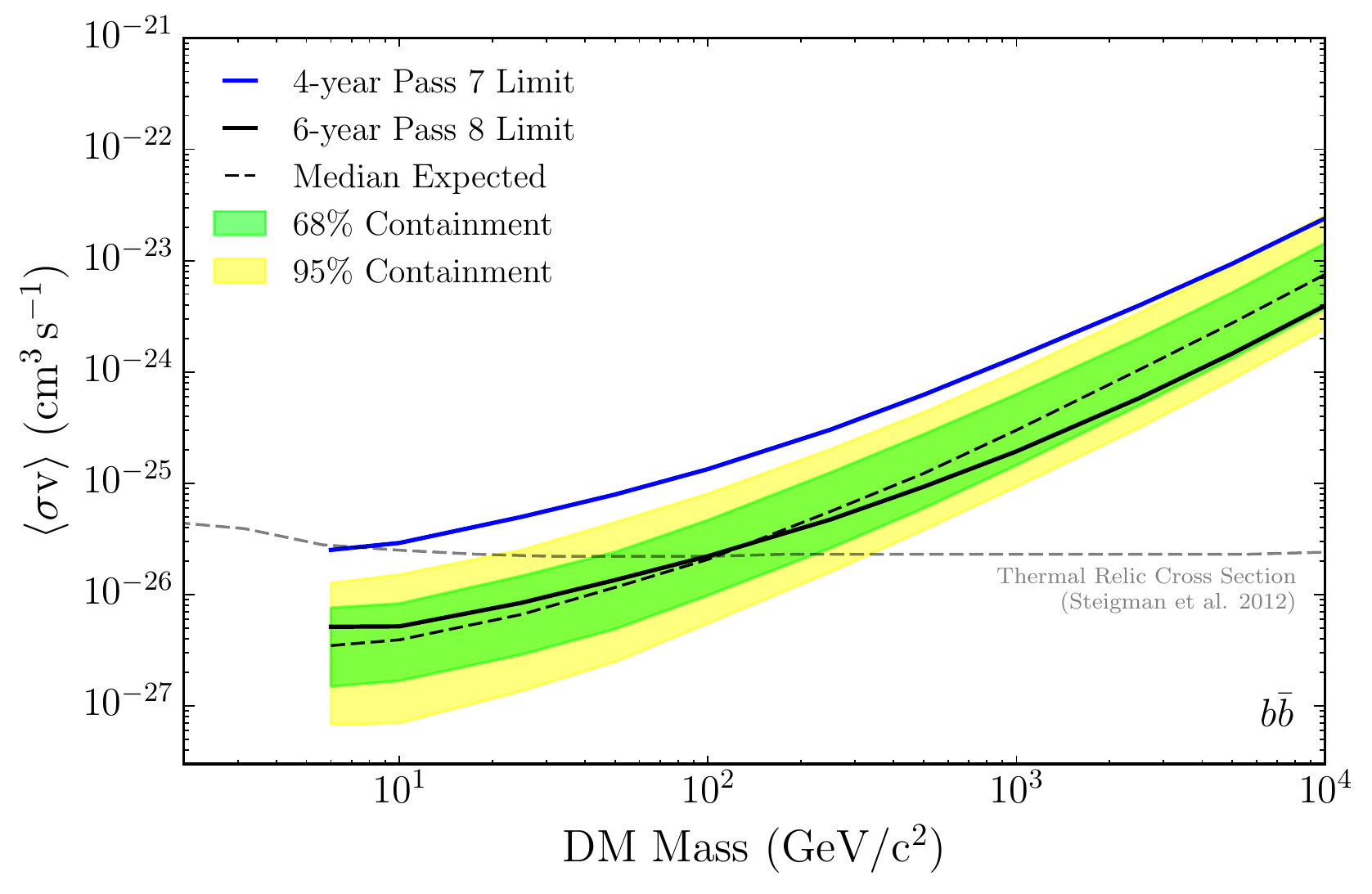}
  \includegraphics[width=0.49\textwidth]{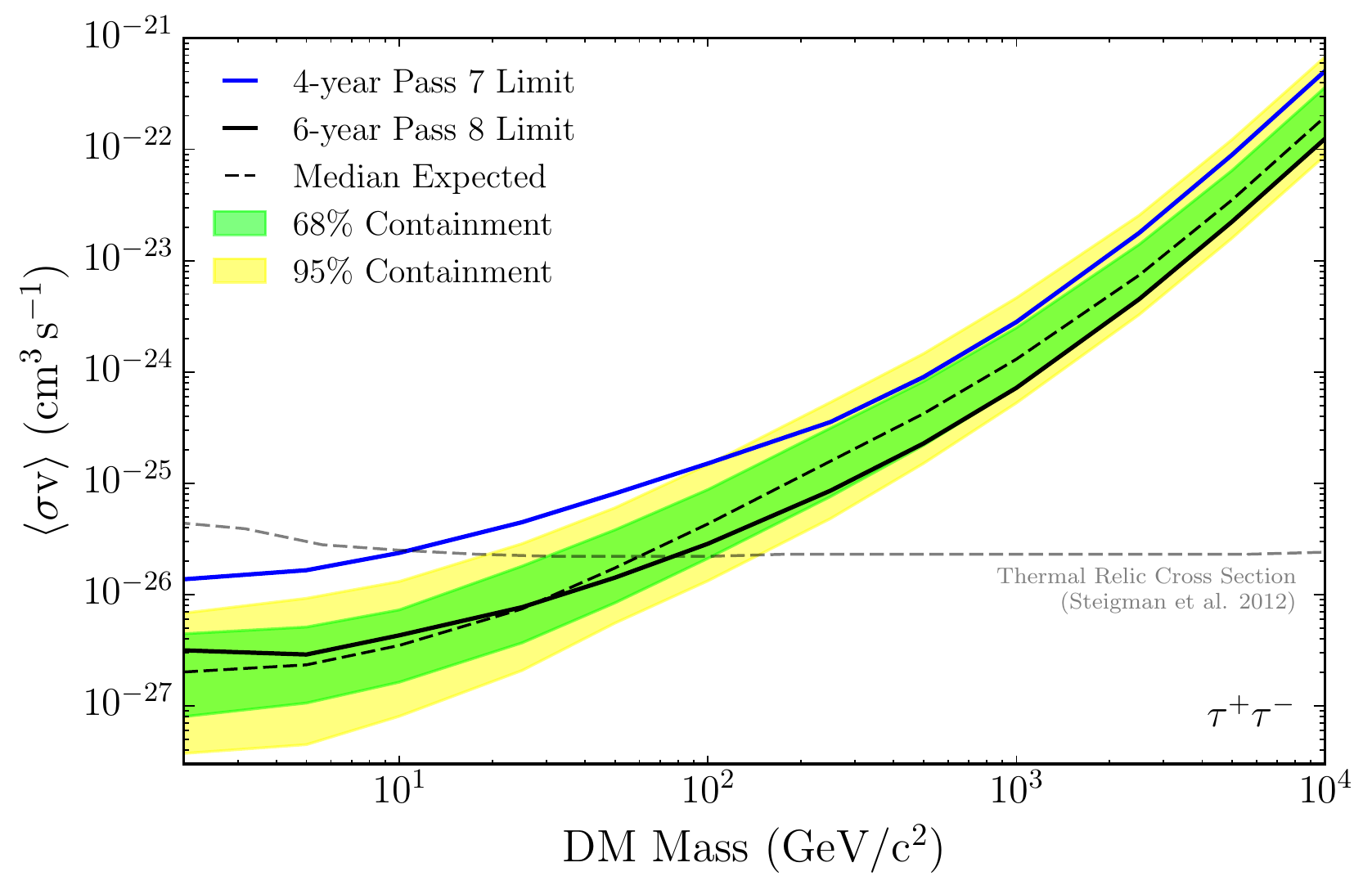}
  \caption{Constraints on the DM annihilation cross section at 95\%
    \CL for the \bbbar (left) and \tautau (right) channels derived
    from a combined analysis of 15 dSphs.  Bands for the expected
    sensitivity are calculated by repeating the same analysis on
    \NRAND randomly selected sets of high-Galactic-latitude blank
    fields in the LAT data.  The dashed line shows the median expected
    sensitivity while the bands represent the 68\% and 95\% quantiles.
    For each set of random locations, nominal \Jfactors are randomized
    in accord with their measurement uncertainties.  The solid blue
    curve shows the limits derived from a previous analysis of four
    years of \passsevenrep data and the same sample of 15 dSphs
    \cite{Ackermann:2013yva}.  The dashed gray curve corresponds to
    the thermal relic cross section from
    \cite{Steigman:2012nb}.}\label{fig:fig1}
\end{figure*}

We find no evidence for significant gamma-ray emission from any of the
individual dSphs or in the combined sample.  Using the joint
likelihood for the 15 dSphs in the combined sample, we derive limits
on the WIMP annihilation cross section as a function of WIMP mass.
Figure \ref{fig:fig1} shows the combined limits evaluated for the
\bbbar and \tautau annihilation channels.  The expectation bands show the expected
range for these limits based on an analysis of 300 sets of
randomly selected blank fields with $|b| > 30\degree$.  We find that the limits from the
six-year \passeight analysis improve on the limits of
\cite{Ackermann:2013yva} by a factor between 3 and 5.  This
improvement can be attributed to a number of factors including the
improved sensitivity of the \passeight data set and the different
statistical realizations of the two data sets.  Because the \passeight
six-year and \passsevenrep four-year event samples have a shared
fraction of only 20--40\%, the two analyses are nearly statistically
independent.  For masses below 100\GeV, the upper limits of
\cite{Ackermann:2013yva} were near the 95\% upper bound of the
expected sensitivity band while the limits in the present analysis are
within one standard deviation of the median expectation value.

\begin{figure*}[ht]
  \centering
  \includegraphics[width=0.49\textwidth]{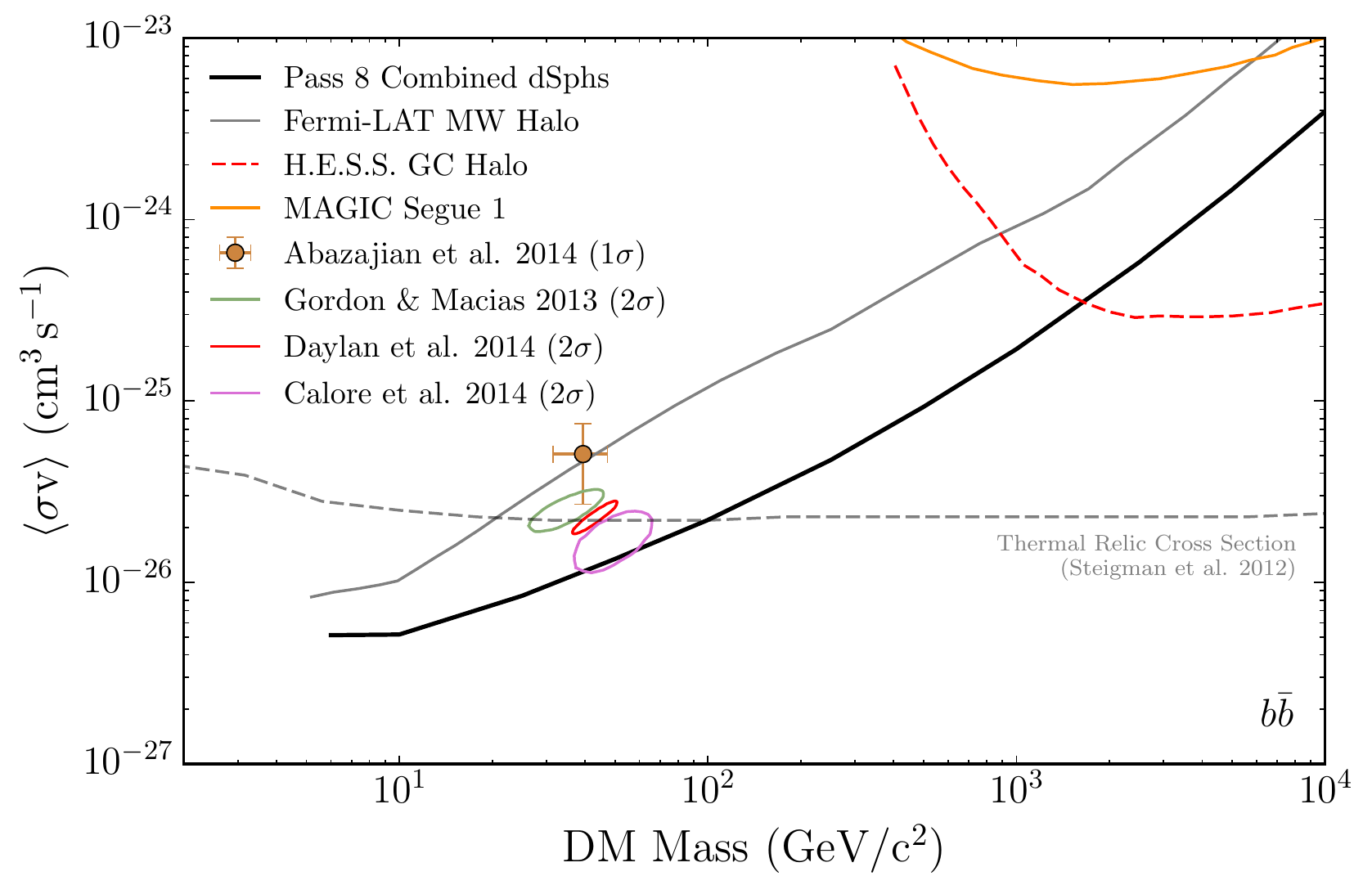}
  \includegraphics[width=0.49\textwidth]{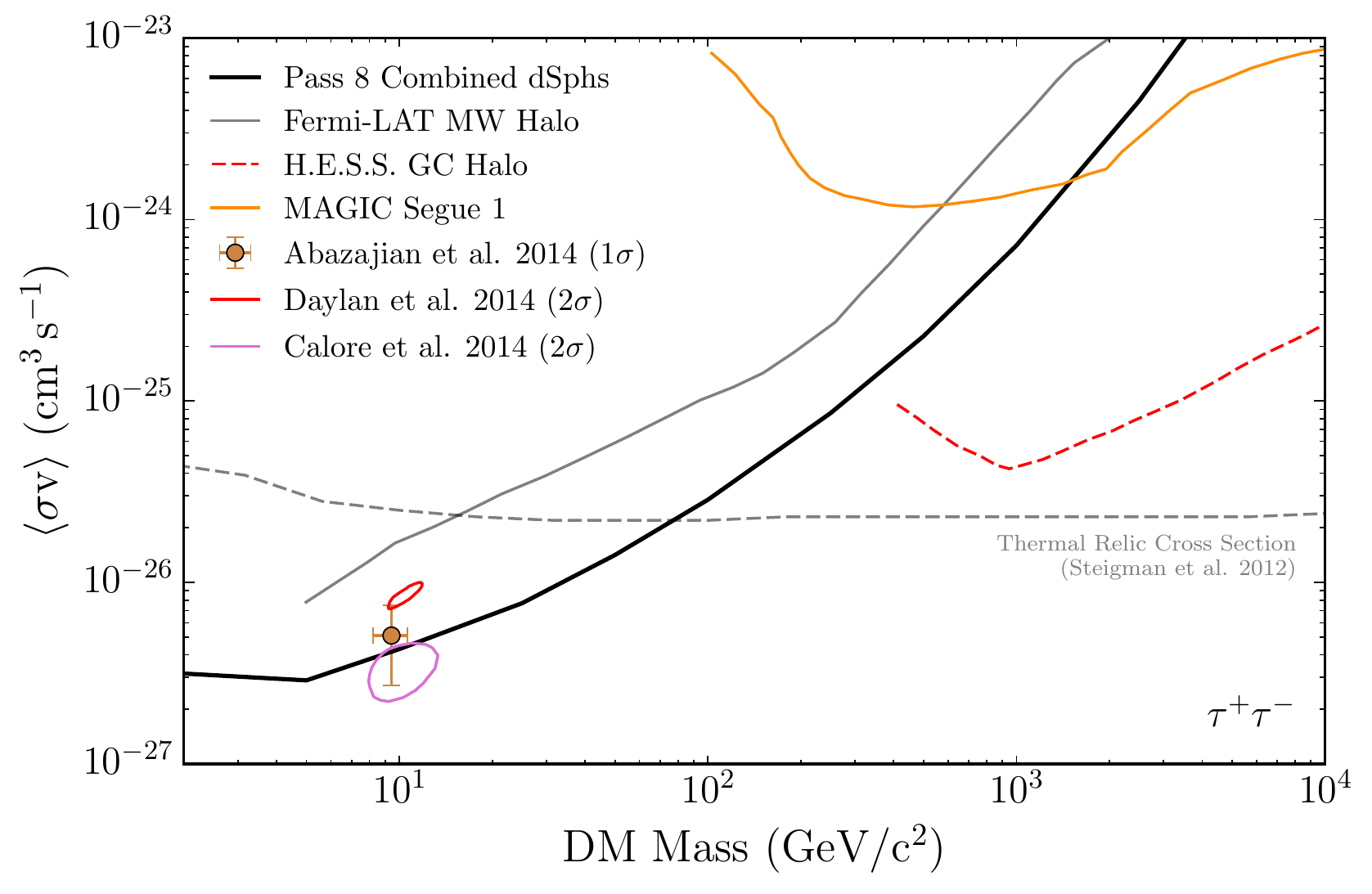}
  \caption{Constraints on the DM annihilation cross at 95\% \CL
    section for the \bbbar (left) and \tautau (right) channels derived
    from the combined analysis of 15 dSphs with 6 years of Pass 8
    data.  For comparison limits from previously published searches
    are shown from LAT analysis of the Milky Way halo ($3\sigma$
    limit) \cite{Ackermann:2012rg}, 112 hours of observations of the
    Galactic Center with H.E.S.S.  \cite{Abramowski:2011hc}, and 157.9
    hours of observations of {Segue~1} with MAGIC
    \cite{Aleksic:2013xea}.  Pure annihilation channel limits for the
    Galactic Center H.E.S.S. observations are taken from
    \cite{Abazajian:2011ak} and assume an Einasto Milky Way density
    profile with $\rho_\odot = 0.389$\GeVcmcube.  Closed contours and
    the marker with error bars show the best-fit cross section and
    mass from several interpretations of the Galactic center excess
    \cite{Abazajian:2014fta,Calore:2014xka,Gordon:2013vta,Daylan:2014rsa}.}\label{fig:fig2}
\end{figure*}

Figure \ref{fig:fig2} shows the comparison of the limits from this
work with other published limits on the DM annihilation cross section.
The \passeight combined dSph limits are currently among the strongest
experimental constraints on the WIMP annihilation cross section and
exclude WIMPs with the thermal relic cross section for ${\mDM \lesssim
  100\GeV}$.  Our results also constrain DM particles with \mDM above
100\GeV surpassing the best limits from Imaging Atmospheric Cherenkov
Telescopes for masses up to $\roughly 1\TeV$ for quark channels and
$\roughly 300\GeV$ for the $\tau$-lepton channel.

These limits also begin to constrain the preferred region of parameter
space for DM interpretations of the Galactic Center
excess~\cite{Gordon:2013vta,Abazajian:2014fta,Daylan:2014rsa,Calore:2014xka}.
However, uncertainties in the Galactic DM
distribution can significantly enlarge the best-fit regions of
\sigmav, channel, and \mDM \cite{2014arXiv1411.2592A}.

The future sensitivity to DM annihilation in dSphs will benefit from
additional LAT data taking and the discovery of new dSphs with current
and upcoming optical surveys such as the Dark Energy Survey
\cite{Abbott:2005bi} and the Large Synoptic Survey Telescope
\cite{Ivezic:2008fe}.  The recent discovery of new dSph candidates in
the first year of the DES survey
\cite{Bechtol:2015cbp,Koposov:2015cua} suggests that the sample of
dSphs that can be targeted for gamma-ray searches will continue to
grow over the next 5--10 years.  Future analyses incorporating these
new dSph galaxies will improve the sensitivity of the LAT to a DM
signal and allow the DM annihilation parameter space to be probed to
even lower cross sections.

\acknowledgments

The \textit{Fermi}-LAT Collaboration acknowledges support for LAT
development, operation and data analysis from NASA and DOE (United
States), CEA/Irfu and IN2P3/CNRS (France), ASI and INFN (Italy), MEXT,
KEK, and JAXA (Japan), and the K.A.~Wallenberg Foundation, the Swedish
Research Council and the National Space Board (Sweden). Science
analysis support in the operations phase from INAF (Italy) and CNES
(France) is also gratefully acknowledged.

\bibliographystyle{JHEP}
\bibliography{bib}

\providecommand{\href}[2]{#2}\begingroup\raggedright\begin{thebibliography}{10}

\bibitem{Ackermann:2013yva}
{\bf Fermi-LAT Collaboration} Collaboration, M.~Ackermann et~al., {\it {Dark
  Matter Constraints from Observations of 25 Milky Way Satellite Galaxies with
  the Fermi Large Area Telescope}},  {\em \prd} {\bf 89} (2014) 042001,
  [\href{http://arxiv.org/abs/1310.0828}{{\tt arXiv:1310.0828}}].

\bibitem{Ackermann:2015zua}
{\bf Fermi-LAT} Collaboration, M.~Ackermann et~al., {\it {Searching for Dark
  Matter Annihilation from Milky Way Dwarf Spheroidal Galaxies with Six Years
  of Fermi-LAT Data}},  \href{http://arxiv.org/abs/1503.0264}{{\tt
  arXiv:1503.0264}}.

\bibitem{Ackermann:2011wa}
{\bf Fermi-LAT Collaboration} Collaboration, M.~Ackermann et~al., {\it
  {Constraining Dark Matter Models from a Combined Analysis of Milky Way
  Satellites with the Fermi Large Area Telescope}},  {\em \prl} {\bf 107}
  (2011) 241302, [\href{http://arxiv.org/abs/1108.3546}{{\tt
  arXiv:1108.3546}}].

\bibitem{GeringerSameth:2011iw}
A.~Geringer-Sameth and S.~M. Koushiappas, {\it {Exclusion of canonical WIMPs by
  the joint analysis of Milky Way dwarfs with Fermi}},  {\em \prl} {\bf 107}
  (2011) 241303, [\href{http://arxiv.org/abs/1108.2914}{{\tt
  arXiv:1108.2914}}].

\bibitem{Mazziotta:2012ux}
M.~N. Mazziotta, F.~Loparco, F.~de~Palma, and N.~Giglietto, {\it {A
  model-independent analysis of the Fermi Large Area Telescope gamma-ray data
  from the Milky Way dwarf galaxies and halo to constrain dark matter
  scenarios}},  {\em Astropart.} {\bf 37} (2012) 26--39,
  [\href{http://arxiv.org/abs/1203.6731}{{\tt arXiv:1203.6731}}].

\bibitem{Geringer-Sameth:2014qqa}
A.~Geringer-Sameth, S.~M. Koushiappas, and M.~G. Walker, {\it {A Comprehensive
  Search for Dark Matter Annihilation in Dwarf Galaxies}},
  \href{http://arxiv.org/abs/1410.2242}{{\tt arXiv:1410.2242}}.

\bibitem{2015arXiv150203081A}
B.~{Anderson}, J.~{Chiang}, J.~{Cohen-Tanugi}, J.~{Conrad}, A.~{Drlica-Wagner},
  M.~{Llena Garde}, and {Stephan Zimmer for the Fermi-LAT Collaboration}, {\it
  {Using Likelihood for Combined Data Set Analysis}},  {\em ArXiv e-prints}
  (Feb., 2015) [\href{http://arxiv.org/abs/1502.0308}{{\tt arXiv:1502.0308}}].

\bibitem{Atwood:2013rka}
{\bf Fermi-LAT Collaboration} Collaboration, W.~Atwood et~al., {\it {Pass 8:
  Toward the Full Realization of the Fermi-LAT Scientific Potential}},  {\em
  eConf} {\bf C121028} (2013) [\href{http://arxiv.org/abs/1303.3514}{{\tt
  arXiv:1303.3514}}].

\bibitem{Ackermann:2015hja}
{\bf The Fermi-LAT Collaboration} Collaboration, M.~Ackermann et~al., {\it
  {Fermi Large Area Telescope Third Source Catalog}},
  \href{http://arxiv.org/abs/1501.0200}{{\tt arXiv:1501.0200}}.

\bibitem{Martinez:2013ioa}
G.~D. Martinez, {\it {A Robust Determination of Milky Way Satellite Properties
  using Hierarchical Mass Modeling}},  {\em \mnras} {\bf 451} (2015)
  2524--2535, [\href{http://arxiv.org/abs/1309.2641}{{\tt arXiv:1309.2641}}].

\bibitem{Rolke:2004mj}
W.~A. Rolke, A.~M. Lopez, and J.~Conrad, {\it {Limits and confidence intervals
  in the presence of nuisance parameters}},  {\em Nucl.Instrum.Meth.} {\bf
  A551} (2005) 493--503, [\href{http://arxiv.org/abs/physics/0403059}{{\tt
  physics/0403059}}].

\bibitem{Steigman:2012nb}
G.~Steigman, B.~Dasgupta, and J.~F. Beacom, {\it {Precise Relic WIMP Abundance
  and its Impact on Searches for Dark Matter Annihilation}},  {\em \prd} {\bf
  86} (2012) 023506, [\href{http://arxiv.org/abs/1204.3622}{{\tt
  arXiv:1204.3622}}].

\bibitem{Ackermann:2012rg}
{\bf Fermi-LAT Collaboration} Collaboration, M.~Ackermann et~al., {\it
  {Constraints on the Galactic Halo Dark Matter from Fermi-LAT Diffuse
  Measurements}},  {\em \apj} {\bf 761} (2012) 91,
  [\href{http://arxiv.org/abs/1205.6474}{{\tt arXiv:1205.6474}}].

\bibitem{Abramowski:2011hc}
{\bf H.E.S.S.~Collaboration} Collaboration, A.~Abramowski et~al., {\it {Search
  for a Dark Matter annihilation signal from the Galactic Center halo with
  H.E.S.S}},  {\em \prl} {\bf 106} (2011) 161301,
  [\href{http://arxiv.org/abs/1103.3266}{{\tt arXiv:1103.3266}}].

\bibitem{Aleksic:2013xea}
{\bf MAGIC Collaboration} Collaboration, J.~Aleksi\'{c}, S.~Ansoldi,
  L.~Antonelli, P.~Antoranz, A.~Babic, et~al., {\it {Optimized dark matter
  searches in deep observations of Segue 1 with MAGIC}},  {\em \jcap} {\bf
  1402} (2014) 008, [\href{http://arxiv.org/abs/1312.1535}{{\tt
  arXiv:1312.1535}}].

\bibitem{Abazajian:2011ak}
K.~N. Abazajian and J.~P. Harding, {\it {Constraints on WIMP and
  Sommerfeld-Enhanced Dark Matter Annihilation from HESS Observations of the
  Galactic Center}},  {\em JCAP} {\bf 1201} (2012) 041,
  [\href{http://arxiv.org/abs/1110.6151}{{\tt arXiv:1110.6151}}].

\bibitem{Abazajian:2014fta}
K.~N. Abazajian, N.~Canac, S.~Horiuchi, and M.~Kaplinghat, {\it {Astrophysical
  and Dark Matter Interpretations of Extended Gamma-Ray Emission from the
  Galactic Center}},  {\em \prd} {\bf 90} (2014) 023526,
  [\href{http://arxiv.org/abs/1402.4090}{{\tt arXiv:1402.4090}}].

\bibitem{Calore:2014xka}
F.~Calore, I.~Cholis, and C.~Weniger, {\it {Background model systematics for
  the Fermi GeV excess}},  \href{http://arxiv.org/abs/1409.0042}{{\tt
  arXiv:1409.0042}}.

\bibitem{Gordon:2013vta}
C.~Gordon and O.~Macias, {\it {Dark Matter and Pulsar Model Constraints from
  Galactic Center Fermi-LAT Gamma Ray Observations}},  {\em \prd} {\bf 88}
  (2013), no.~4 083521, [\href{http://arxiv.org/abs/1306.5725}{{\tt
  arXiv:1306.5725}}].

\bibitem{Daylan:2014rsa}
T.~Daylan, D.~P. Finkbeiner, D.~Hooper, T.~Linden, S.~K.~N. Portillo, et~al.,
  {\it {The Characterization of the Gamma-Ray Signal from the Central Milky
  Way: A Compelling Case for Annihilating Dark Matter}},
  \href{http://arxiv.org/abs/1402.6703}{{\tt arXiv:1402.6703}}.

\bibitem{2014arXiv1411.2592A}
P.~{Agrawal}, B.~{Batell}, P.~J. {Fox}, and R.~{Harnik}, {\it {WIMPs at the
  Galactic Center}},  {\em ArXiv e-prints} (Nov., 2014)
  [\href{http://arxiv.org/abs/1411.2592}{{\tt arXiv:1411.2592}}].

\bibitem{Abbott:2005bi}
{\bf DES Collaboration} Collaboration, T.~Abbott et~al., {\it {The dark energy
  survey}},  \href{http://arxiv.org/abs/astro-ph/0510346}{{\tt
  astro-ph/0510346}}.

\bibitem{Ivezic:2008fe}
{\bf LSST Collaboration} Collaboration, Z.~Ivezic, J.~Tyson, R.~Allsman,
  J.~Andrew, and R.~Angel, {\it {LSST: from Science Drivers to Reference Design
  and Anticipated Data Products}},  \href{http://arxiv.org/abs/0805.2366}{{\tt
  arXiv:0805.2366}}.

\bibitem{Bechtol:2015cbp}
{\bf DES} Collaboration, K.~Bechtol et~al., {\it {Eight New Milky Way
  Companions Discovered in First-Year Dark Energy Survey Data}},
  \href{http://arxiv.org/abs/1503.0258}{{\tt arXiv:1503.0258}}.

\bibitem{Koposov:2015cua}
S.~E. Koposov, V.~Belokurov, G.~Torrealba, and N.~W. Evans, {\it {Beasts of the
  Southern Wild: Discovery of nine Ultra Faint satellites in the vicinity of
  the Magellanic Clouds}},  {\em Astrophys.J.} {\bf 805} (2015), no.~2 130,
  [\href{http://arxiv.org/abs/1503.0207}{{\tt arXiv:1503.0207}}].

\end{thebibliography}\endgroup


\end{document}